# Friendliness Of Stack Overflow Towards Newbies

Aneesh Tickoo[1], Shweta Chauhan[1], Gagan Raj Gupta[1][gagan@iitbhilai.ac.in]

[1] Indian Institute of Technology, Bhilai CG 492015, INDIA

**Abstract.** In today's modern digital world, we have a number of online Question and Answer platforms like Stack Exchange, Quora, and GFG that serve as a medium for people to communicate and help each other. In this paper, we analyzed the effectiveness of Stack Overflow in helping newbies to programming. Every user on this platform goes through a journey. For the first 12 months we consider them to be a newbie. Post 12 months they come under one of the following categories: *Experienced*, *Lurkers* or *Inquisitive*. Each question asked has tags assigned to it and we observe that questions with some specific tags have a faster response time indicating an active community in that field over others. The platform had a steady growth up to 2013 after which it started declining, but recently during the pandemic 2020, we can see rejuvenated activity on the platform.

.

## 1     Introduction

Online communities can be a medium to bring together users with similar ideologies in one place where they can discuss course content, apply their learning, and ask questions. Thus, it becomes important to attract new users who are interested in the goal of their community. At the same time, retaining the previous users is also important because they are expected to be more experienced and can help newbies to the community.

Content from these platforms spans numerous spheres of life — finance, things related to business dealings, software development, academics, etc. We refer to these platforms as content-based platforms and study their successes and failures.[1] The addition of new users depends on how user-friendly or newbie-friendly that platform actually is. Because the more user-friendly it is, the more is the number of users that will connect to it and help the community on a frequent basis.

This paper aims to analyze how newbie-friendly these community question-answer platforms are. In particular, we will consider Stack Overflow which is a question-and-answer website for professional and enthusiastic programmers. Stack Overflow is part of the Stack Exchange Network which comprises many other sites. It is established that users in these communities perform diverse and essential roles.[2] It becomes important to recognize these roles and how users transition between them. Many previous works have tried to categorize users into various roles.[3]

Existing techniques fall short of our objectives in a variety of ways. Many existing models don't consider the timeline of a user rather work on the current snapshot of data. We have extracted the first 12 months' activity of each user and compared it to the user's post 12-month time spent on the platform. In a way we are analyzing each user's journey on the platform as he transforms from a newbie to an older member.

We also aim to study if older members of the community are still active on these platforms or not. The platform assigns a reputation to each user which indicates the value that you have added to the platform. It's a good measure to separate the user base into various user types. On analysis, we find out 66.14% of the user base have no activity on the platform we term them *Silent Observers*. A user is considered a *Newbie* for the first 12 months from their joining. We also categorized users as *Experienced* (Reputation>200 and at least 5 answers on the platform), *Lurkers* (Reduced activity post their 12-month mark), *Inquisitive* (Activity even after 12 months but majority of the posts are Questions).

We have analyzed the trends of questions asked and answered from the very start of the platform in 2008 to the current day with a special emphasis on the period when education was affected by COVID-19.[4] Apart from this, we will study the response times for the questions asked on this platform. We will also try to uncover the aspects that cause longer response times.

Fig 1 shows a sample of the XML file used in data processing.

```xml
<?xml version="1.0" encoding="utf-8"?>
<posts>
  <row Id="5" PostTypeId="1" CreationDate="2014-05-13T23:58:30.457" Score="9" ViewCount="799" Body="<
  <row Id="7" PostTypeId="1" AcceptedAnswerId="10" CreationDate="2014-05-14T00:11:06.457" Score="4" Viev
  <row Id="9" PostTypeId="2" ParentId="5" CreationDate="2014-05-14T00:36:31.077" Score="5" Body="<p&
  <row Id="10" PostTypeId="2" ParentId="7" CreationDate="2014-05-14T00:53:43.273" Score="13" Body="<p
  <row Id="14" PostTypeId="1" AcceptedAnswerId="29" CreationDate="2014-05-14T01:25:59.677" Score="25" V
  <row Id="15" PostTypeId="1" CreationDate="2014-05-14T01:41:23.110" Score="2" ViewCount="648" Body="<
  <row Id="16" PostTypeId="1" AcceptedAnswerId="46" CreationDate="2014-05-14T01:57:56.880" Score="17" V
  <row Id="17" PostTypeId="5" CreationDate="2014-05-14T02:49:14.580" Score="0" Body="<p><a hret
```

**Fig. 1** Snapshot of the sample data of Posts.xml

## 2    Research Methodology

To begin with our analysis, we acquired the data from the Stack Exchange Data Dump [5] which is an anonymized dump of all user-contributed content on the Stack Exchange network. The data is available in XML format.

Posts.xml has information regarding all the posts of Stack Overflow. The uncompressed size of the data used is close to 100 GB and so to analyze this we have used Map-Reduce approach in Apache Hadoop using a 5 Node AWS EMR cluster. Schema of Stack Overflow XML files as shown in Table 2. The rows have the names of the different XML files available and the columns denote the XML tags in the respective files. Our fundamental analysis involves Posts.xml and Users.xml. The basic statistics of the Stack Overflow platform are as shown in Table 1.

**Table 1.** Basic statistics of the Stack Overflow platform

| | |
|---|---|
| Total Number of Users | 16,279,651 |
| Average Reputation[1] of a User | 106 |
| Total Number of Questions Asked | 21,978,326 |
| Total Number of Answers posted | 20,815,178 |
| Total Questions with an accepted answer | 10,682,706 |
| % Of accepted answers[2] | 51.39 % |
| Average Response Time (in hours) | 252.7 |

**Table 1.** Schema of Stack Overflow XML files

| | User Id | Post Id | Post Type Id | Creation Date | Reputation | View Count | Tags |
|---|---|---|---|---|---|---|---|
| Posts | Yes | - | Yes | Yes | - | Yes | Yes |
| Users | Yes | - | - | Yes | Yes | - | - |
| Badges | Yes | - | - | - | - | - | - |
| Comments | Yes | Yes | - | Yes | - | - | - |
| Post history | Yes | Yes | - | Yes | - | - | - |
| Post links | - | Yes | - | Yes | - | - | - |
| Tags | - | - | - | - | - | - | Yes |
| Votes | - | Yes | - | Yes | - | - | - |

To analyze the Stack Exchange dataset, we use the Map-Reduce approach as it is scalable across many nodes. [6]

The MapReduce algorithm contains two important tasks, namely Map and Reduce. [7]

- The Map task takes a set of data and converts it into another set of data, where individual elements are broken down into tuples (key-value pairs).

- The Reduce task takes the output from the Map as an input and combines those data tuples (key-value pairs) into a smaller set of tuples.

The reduce task is always performed after the map task. These tasks can be done over distributed systems on chunks of our input data allowing faster processing of big data.

```python
#!/usr/bin/env python
# import sys because we need to read and write data to STDIN and STDOUT
import sys
import re        #ReGex was used to match the required data field

user_id = re.compile(r'row Id="(\d*)"')
reputation = re.compile(r'Reputation="(\d*)"')
for line in sys.stdin:
    try:
        uid = user_id.search(line)
        rep = reputation.search(line)
        print("{}\t{}".format(uid.group(1),rep.group(1)))
        # outputs (User_Id , Reputation)
    except:
        pass
```

**Fig. 2.** shows a snapshot of the mapper code

The mapper code uses regular expressions to extract the reputation of the users from the acquired dataset. The final output tuple is (User_Id, reputation).

```python
#!/usr/bin/env python
# import sys because we need to read and write data to STDIN and STDOUT
import sys
current_rep=""
count=0
for line in sys.stdin:
    line = line.strip()
    #streaming input received from mapper's output
    uid,rep = line.split("\t", 1)
    if rep==current_rep:
        count+=1
    else:
        if current_rep:
            print("{}\t{}".format(current_rep,count))
            count=0
        current_rep=rep
        count+=1
#Important to take care of last value
print("{}\t{}".format(current_rep,count))
```

**Fig. 3.** shows a snapshot of the reducer code

The reducer code takes the output of the mapper code as its input and produces a tuple containing the reputation and the number of users having that aforementioned reputation.

## 3       Key Findings

### 3.1    User Type Analysis

On Stack Overflow, each user's time on the platform can be visualized broadly into 5 categories on the basis of their activity in terms of questions asked, answers given, and questions resolved before and after the 12-month mark. The categories are defined as follows:

1. *The Silent Observer*: 66.14% of users fall in this category wherein no questions or answers are posted ever.
2. *Experienced*: 2.45 % of users fall into this category where they have answered a number of questions and have significantly contributed to the community.
3. *Inquisitive:* 6.77% of users fall under this category wherein they continue to post questions even after the 12-month mark but do not make significant contributions in terms of answers.
4. *Lurker*: These 23.34% are the users who have very low frequency of contributions to the community in terms of both questions and answers, signifying that they have lost interest.
5. *Others*: This 1.28% of the users do not fall in any of the above categories. The majority of these users are those who have high reputations but have posted fewer answers. The reason behind their high reputation is the fact that they have noteworthy contributions on other SE websites which is another criterion taken into consideration during reputation calculation by SE.

Table 3 and Fig 4 summaries the above details.

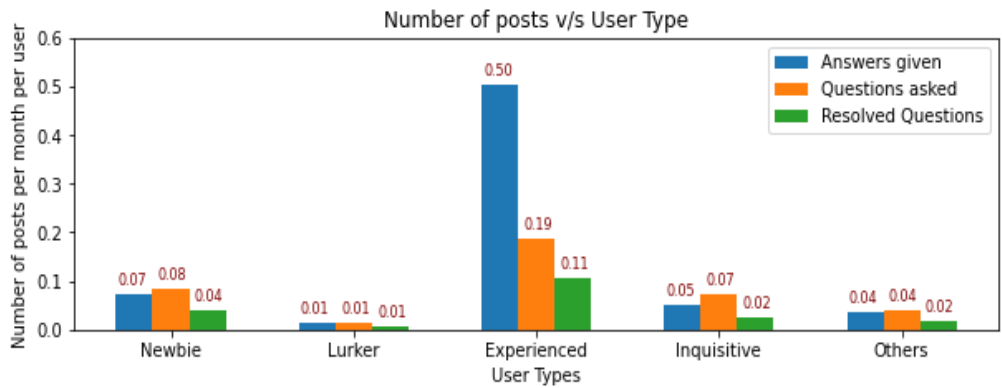

**Fig. 4.** is a graph plotted between the type of user and the number of questions they have asked or answered. The blue bars show the number of answers given by the type of user, the orange ones show the number of questions asked and the green bars show the number of resolved questions. This graph shows a comparison between the users' first 12 months on the platform [*Newbie*] VS their transformation post-12-months of their joining. [*Lurker*, *Experienced*, *Inquisitive*, *Others*].

**Table 3.** Distribution of Users in various categories

| | | |
|---|---|---|
| Total Users | 16,279,651 | |
| Silent Observer | 10,768,975 | **66.14%** |
| Experienced | 398,960 | **2.45%** |
| Inexperienced but interested | 1,102,495 | **6.77%** |
| Lurker | 3,799,964 | **23.34%** |
| Others | 209,252 | **1.28%** |

## 3.2 Response Time analysis

Newbies are new to the technological field and seek answers to their questions. These are provided by the more experienced users on the platform and help the community as a whole. This difference also shows the credibility of the reputation system that Stack Exchange uses as the user reputation changes with time and a user with a higher reputation is likely to be more experienced in a specified field.[8]

The average response time on Stack Exchange is 252.7 hours. For plotting Fig 5, we have considered some tags that became popular, that is, had a large number of questions asked related to them. As can be seen, they have slower response times. Lower response

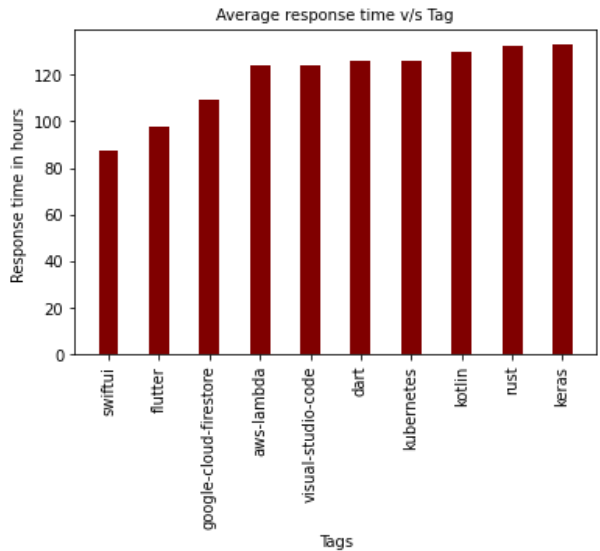

**Fig. 5.** shows the average time taken to get an adequate response for a question under the specified tag. A lesser response time means an active community exists for that tag

time indicates that many users are knowledgeable in the related field and thus leads us to believe that these topics either recently became relevant or have continued to maintain their importance in the programming field.

As most of these tags are related to relatively new software and app development frameworks, we can conclude that they are currently popular in this field. More and more people are becoming interested in these topics which has led to such a high number of questions being answered in short spans of time. Also, often the answers posted on the platform point to various other websites that are beneficial for the topics. We have extracted website links from each answer and compiled the result.

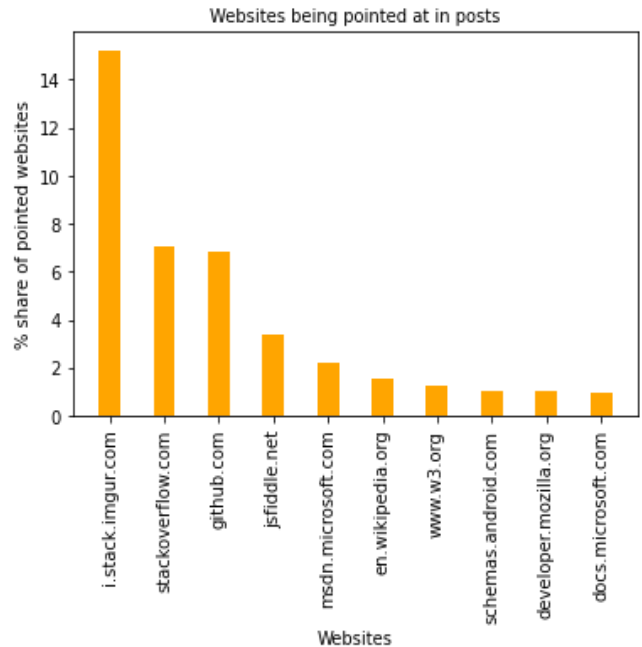

**Fig. 6.** shows the websites which have been referenced most by the users on the Stack Exchange community in their posts.

In Fig 6 we can see that the most pointed websites came out to be Imgur, Stack Overflow, GitHub, JSFiddle, MSDN from Microsoft, and many others related to the field of software development. Imgur came out to be the most pointed website. It is an image hosting and sharing site which is favored by users of social media and other CQA (common question-answering) communities to share images with their posts.

Stack Overflow was the second most pointed website because many users refer to the answers to the questions previously posted by other users. Content is heavily curated by the community of Stack Overflow. Duplicate or similar questions are quickly identified as such and merged with existing questions. Most of the comments on Stack Exchange pointed to repository hosting services like GitHub, which provides Internet hosting for software development offering distributed version control and source code management functionality. The next one in the popularity line is JSFiddle, which is an online IDE service and it is also an online community for testing and showcasing user-created HTML, CSS, and JavaScript code snippets, known as 'fiddles'. MSDN is also on the list because it allows users to install and use software to design, develop, test, evaluate, and demonstrate other software. The above-mentioned websites like Imgur, JSFiddle, MSDN Microsoft play the role of supporting websites and they are not knowledge-based websites because they don't provide answers.

These results were evident from the fact that Stack Overflow primarily focuses on questions about programming that is tightly focused on a specific problem. This gives us a clear insight into how the community chooses to share and convey their questions or answers.

### 3.3 Activity Analysis during COVID

As of this writing, the new coronavirus infection (COVID-19), which was designated a pandemic by the World Health Organization (WHO) on March 11, 2020, had infected over 271 million individuals worldwide and killed over 5.33 million people. The worldwide disaster caused by the COVID-19 outbreak has driven the adoption of technical innovations and digital goods with the objective of lowering or at the very least mitigating the pandemic's short and/or long-term consequences.

COVID-19 has led in a significant growth in Software Development, as well as a rapid demand for digitalized platforms that are extending the transformation of homes into destinations for remote education and employment. Despite the importance of software development in research, it has been said that domain scientists are under-trained in software development. As a result, we suggest that domain scientists should use knowledge-sharing forums on a regular basis to solve development problems. [9]

The major style of teaching was moved to an online or hybrid approach during the epidemic. Thus, students were attracted to various online learning platforms to improve/build their skills. We have tried to analyze the activity on Stack Overflow from its very start to date. The data ranges from July 2008 to November 2021. We plotted the number of questions asked and the posts answered every month starting from July 2008 all the way up till November 2021.

The user activity increased by leaps and bounds during the peak of the COVID-19 period. This makes it evident that since everyone was locked up in their homes, users became more and more active thus leading to such jumps in the activity graph. After the COVID-19 period, we can clearly see that the difference between the number of questions being asked and those being answered is diminishing; this indicates the reduced activity of users.

On a general level, we can see the platform activity peaked in 2013 and since then it's decreasing with each passing year with a slight peak in 2020. A possible reason for the decrease in the number of posts might be because Stack Overflow has now become a good repository of questions and answers. Many a time the user doesn't need to ask any questions as they have been asked and answered previously on the forum

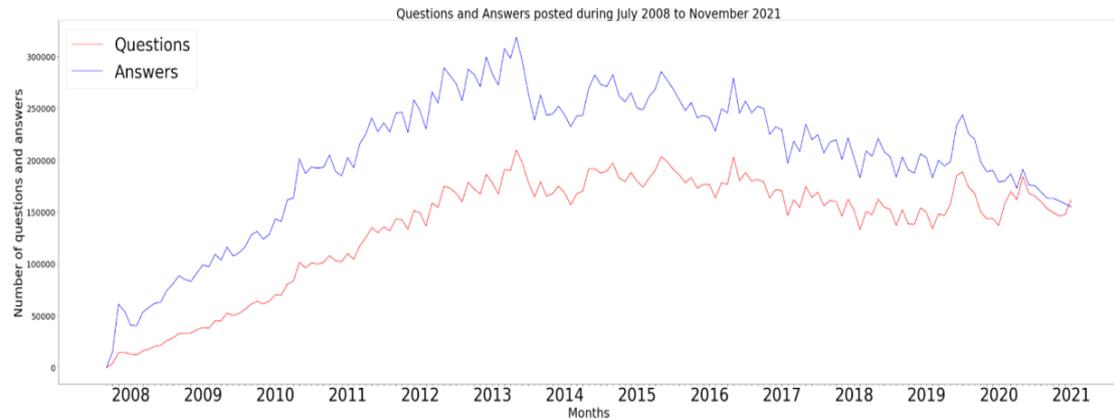

**Fig. 7.** This plot shows the number of questions and answers posted every month from 2008 to 2021. The platform started in July 2008 with only 2 questions and 4 answers. In the first month of January 2014, the platform activity was at its peak with 30,000 questions and 20,000 answers. And ever since the user activity has started to decline with an exception of April to July 2020 (the peak of COVID-19 pandemic). Possible reason of decrease in the number of posts might be because Stack Overflow has now become a good repository of questions and answers. Many a times the user doesn't need to ask any questions as they have been asked and answered previously on the forum.

### 3.4 Question tag analysis

We see that some tags are popular when it comes to response times and questions with these tags get a shorter response time. Here, in this graph Fig 8, we have plotted the number of questions asked under the tag versus the respective tag.

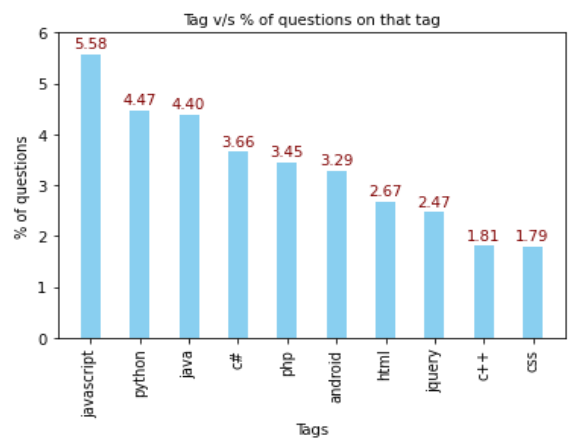

**Fig. 8.** The plot shows the percentage of questions asked of a particular tag

A tag with more questions indicates that more people are trying things related to that topic or are taking interest in the subject. This, by extension, suggests that the topic is very common in the current time and might have a good scope in the future. As we can see, most of the tags shown here are related to software development and programming fields, which suggest that these might be big contributors in shaping the future.

A tag with more accepted answers is likely to be an older topic that has retained the users' interest, which is why more users have useful knowledge about it. To analyze this, we plotted the number of accepted answers against the tags considered previously in Fig 9.

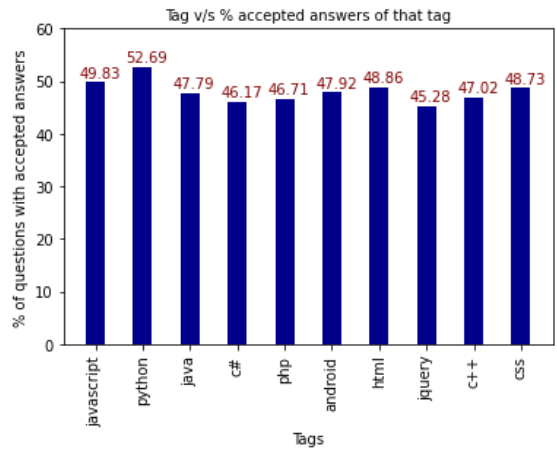

**Fig. 9.** this plot shows the average number of accepted answers for the tags shown in Fig 8.

Going by the statistics we can infer that many websites were made which can be a reason for the increased number of questions with tags like JavaScript, HTML, CSS, etc. Nearly 70% of professional developers who responded to the 2020 Stack Overflow survey coded in JavaScript. JavaScript has been the most-used technology since Stack Overflow started doing the annual survey. [10]

## 4 Discussion

Knowledge-sharing networks based on Q&A are quickly becoming a primary source of information for Internet users. Understanding how users interact on these sites is critical for facilitating information flow between users. [11]

From the above analysis, we can see that Stack Overflow might not constitute of an active community but it does have some experienced users who answer a lot as seen in Fig 4. Since majority of users are *Silent Observers*, this gives us a clear indication that Stack Overflow has developed into a good repository of questions that have been satisfactorily answered. Thus, most of the newer members do not need to ask questions or contribute answers as they get what they desire, serving the objective of the site.

There are certain tags that have a very active community and questions involving those tags get resolved quickly. On a general level, the majority of the questions and answers are based on various programming languages. We can see a decline in the user activity from Fig 7 but it does not clearly imply the platform is dying. A very plausible reason for the decline can be the fact that Stack Overflow has become a huge repository of questions and answers that are still relevant so there are few additions to the already developed question-answer base. Often many users don't need to post a new question as there is already a similar question with a satisfactory answer.

## 5   Conclusion

According to several studies, the vast majority of answers on knowledge-sharing platforms are written by a tiny group of specialists.[12][13] Detecting expert users early on can assist site owners to improve the knowledge produced by their community by highlighting the experts' answers, or it can help them keep these users by providing special privileges. Now regarding whether Stack Exchange is newbie-friendly or not we conclude that it is newbie-friendly but not to a much higher degree. The percentage of accepted answers is around 51.39 %. Out of 6,888,503 answers given by newbies, 4,990,471 answers were accepted by the community which brings us to 72.44 % of accepted answers. Out of 10,682,706 accepted answers, 4,990,471 were given by newbies which brings us around the value of 46.71 %. Thus, we can conclude that a good percentage of answers given by newbies are being accepted by the community and there is not much bias against the new users. New users are definitely benefitting from being part of the community. Community is not declining rather has become a good repository of questions and answers.

## 6   Acknowledgement

We would like to thank Shaleen Malik and Muttareddygari Sreechakra for useful discussions and helpful comments. The authors would like to thank Stack Overflow for providing the data used in this paper.